\documentclass[twocolumn,superscriptaddress]{revtex4-1}
\usepackage{graphicx}
\usepackage[utf8]{inputenc}

\begin{document}

\title{Streak-camera measurements of single photons at telecom wavelength}

\author{Markus Allgaier} \email{markus.allgaier@upb.de}
\affiliation{Integrated Quantum Optics, Applied Physics, University of Paderborn, 33098 Paderborn, Germany}
\author{Vahid Ansari}
\affiliation{Integrated Quantum Optics, Applied Physics, University of Paderborn, 33098 Paderborn, Germany}
\author{Christof Eigner}
\affiliation{Integrated Quantum Optics, Applied Physics, University of Paderborn, 33098 Paderborn, Germany}
\author{Viktor Quiring}
\affiliation{Integrated Quantum Optics, Applied Physics, University of Paderborn, 33098 Paderborn, Germany}
\author{Raimund Ricken}
\affiliation{Integrated Quantum Optics, Applied Physics, University of Paderborn, 33098 Paderborn, Germany}
\author{John Matthew Donohue}
\affiliation{Integrated Quantum Optics, Applied Physics, University of Paderborn, 33098 Paderborn, Germany}
\author{Thomas Czerniuk}
\affiliation{Experimentelle Physik II, Technische Universität Dortmund, D-44221 Dortmund, Germany}
\author{Marc Aßmann}
\affiliation{Experimentelle Physik II, Technische Universität Dortmund, D-44221 Dortmund, Germany}
\author{Manfred Bayer}
\affiliation{Experimentelle Physik II, Technische Universität Dortmund, D-44221 Dortmund, Germany}
\author{Benjamin Brecht}
\affiliation{Integrated Quantum Optics, Applied Physics, University of Paderborn, 33098 Paderborn, Germany}
\affiliation{Clarendon Laboratory, Department of Physics, University of Oxford, Oxford OX1 3PU, United Kingdom}
\author{Christine Silberhorn}
\affiliation{Integrated Quantum Optics, Applied Physics, University of Paderborn, 33098 Paderborn, Germany}

\date{\today}

\begin{abstract}
Up to this point streak-cameras have been a powerful tool for temporal characterization of ultrafast light pulses even at the single photon level. However, the low signal-to-noise ratio in the infrared range prevents measurement on weak light sources in the telecom regime. We present an approach to circumvent this problem. The method utilizes an up-conversion process in periodically poled waveguides in Lithium Niobate. We convert single photons from a parametric down-conversion source in order to reach the point of maximum detection efficiency of commercially available streak-cameras. We explore phase-matching configurations to investigate the up-conversion scheme in real-world applications.
\end{abstract}

\maketitle

\section{Introduction}

Characterization of ultrafast pulses with pulse durations of only a few picoseconds or less is a challenge, especially when it comes to the direct observation in the time domain. Optical autocorrelation measurements \cite{armstrong_measurement_1967} and streak-cameras \cite{campillo_picosecond_1983} are among the most popular techniques to measure the temporal envelope of photons. While indirect temporal measurements using spectral shearing interferometry \cite{fittinghoff_measurement_1996,davis_single-photon_2016}, also known as SPIDER, and frequency resolved optical gating \cite{kane_characterization_1993}, known as FROG, have been demonstrated, streakcameras are able to provide \textit{direct} measurements at the single photon level \cite{wiersig_direct_2009,asmann_ultrafast_2010,asmann_measuring_2010} and have become a standard tool in semiconductor physics. One reason for the device's success is the possibility to record time-resolved spectra in combination with a spectrometer. However, streak-cameras rely on the conversion from light to electrons using photo cathodes, which are inefficient at wavelengths longer than 900\,nm. Early in the development of Streak cameras this issue was circumvented using an up-conversion detection system \cite{onodera_real-time_1983} based on a sum-frequency generation process, but this first implementation suffered from low detection efficiency, limited by the conversion efficiency of the sum-frequency generation. The up-conversion streak-camera was superseded shortly after when better cathode materials made the direct observation of infrared light in the telecom band possible \cite{white_direct_1985}. Using an atomic gas chamber instead of a classical photo cathode improves infrared sensitivity over a broad spectral range \cite{lankhuijzen_far-infrared_1996}, but highlights that further development of telecom-sensitive streak-cameras is technically very challenging. Therefore, up to date there is no streak-camera system capable of detecting telecom light at the single photon level. As streak-cameras have proven to be powerful tools for low-light applications, such a tool for infrared light would certainly proof tremendously useful.

Since the development of the up-conversion streak-camera, there have been more experiments showing the value of up-conversion techniques for single photon measurements. Besides up-conversion assisted photon counting \cite{pelc_long-wavelength-pumped_2011} direct temporal measurements using up-conversion have been demonstrated. One example is the measurement of the temporal intensity of photons from a parametric down-conversion (PDC) source \cite{kuzucu_joint_2008,allgaier_fast_2017}. However, this method cannot provide a simultaneous observation of two degrees of freedom such as time and spectrum, as would be possible with a streak-camera when combined with a spectrometer.

Moreover, many ultrafast phenomena take place on a time scale too short for detection on certain streak-camera models. Especially for low brightness applications there is a trade-off between detection efficiency and spectral resolution. In particular, there is no streak-camera system that can provide both single photon sensitivity in the telecom range and sub picosecond resolution. We propose therefore to not only use an up-conversion process to make light at 1550\,nm accessible to streak-cameras, but also stretch the pulse duration. The concept of pulse stretching while keeping the time-bandwidth-product constant, i.e. bandwidth compression, has been demonstrated using sum-frequency generation \cite{lavoie_spectral_2013}, however, the efficiency of the process needs to be significant to make observation using even the most sensitive streak-cameras possible. We recently introduced a similar scheme for such a process \cite{allgaier_highly_2017}. This process provides both sufficient conversion efficiency and a bandwidth compression factor of 7.5. We demonstrate the performance of the up-conversion device together with a streak-camera by measuring the temporal envelope of single photons from a telecom PDC source. To the best of our knowledge, a measurement of photons from a true single photon source on a streak-camera has not been demonstrated so far, but it is also a rare example of a direct measurements of the temporal envelope of a PDC photon. We study the spectral-temporal properties of the conversion process and their impact on the measurement as well as alternative process conditions in order to provide realistic resolution limits.

\section{Experimental Setup}

\begin{figure}
\centering
\includegraphics[width=0.43\textwidth]{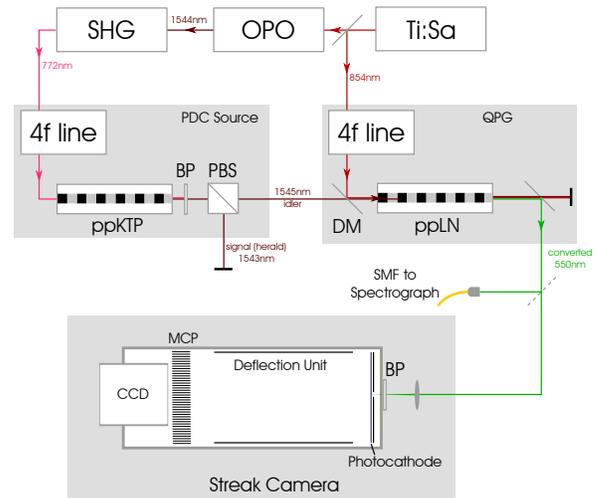}
\caption{Setup used in the experiment. Ti:Sa: Titanium Sapphire laser, OPO: Optical parametric oscillator, SHG: Second harmonic generation, BP: Band pass filter, PBS: Polarizing beam splitter, DM: Dichroic mirror, SMF: Single mode fiber, MCP: Multi channel plate, CCD: Charge coupled device}
\label{fig:setup}
\end{figure}

Figure \ref{fig:setup} depicts our experimental apparatus. It consists of a photon pair source, a frequency converter, and a streak-camera. The photon pair source is a PDC source like the one used in \cite{har2013}. It provides heralded single photons at 1545\,nm central wavelength with a bandwidth of 6\,nm. The source is pumped with pulses from a Ti:Sapphire laser with a pulse energy of 120\,pJ resulting in a moderately low mean photon number of 0.2 per pulse emitted by the source. The heralding efficiency of the source is measured to be 13\,\%. The pump beam for the PDC source is generated by a cascade of an optical parametric oscillator (OPO), and second harmonic generation (SHG) in a periodically poled Lithium Niobate bulk crystal. We use a 4-f bandpass filter to narrow down the pump spectrum. The bandwidth is set to 3\,nm which yields a spectrally decorrelated PDC state \cite{allgaier_highly_2017}.

The pair photons from the type-II process are separated with a polarizing beam splitter, and the heralded photon is sent to the frequency converter. The conversion is achieved by means of a Quantum Pulse Gate (QPG) \cite{eck11,bre2014a}. The device is based on a group-velocity matched sum-frequency generation in Titanium-indiffused waveguides in Lithium Niobate. The process is pumped with pulses with 854\,nm central wavelength from the Ti:Sapphire laser, shaped by a spatial light modulator (SLM) based pulse shaper. The pulse shaper allows us to shape spectral intensity and phase to maximize overlap with the PDC photons and therefore conversion efficiency. Using a 27\,mm long crystal we achieve an internal conversion efficiency of 61.5\,\%. Details regarding the engineering, efficiency, and verification of group-velocity matching of the process are elaborated in \cite{allgaier_highly_2017,ans2016}.

We record the up-converted light at 550\,nm on a Hamamatsu C10910 streak-camera equipped with a S1 photo cathode and ORCA-ER CCD camera. The device's deflection circuit is being operated in the so-called synchroscan mode, where the deflection circuit's repetition rate is synchronized and actively stabilized to the laser's repetition rate of 80.165\,MHz, meaning that every pulse impinges on the CCD. The device's temporal resolution of 5\,ps is mainly limited by the focus spot size on the photo cathode.

\section{Results}

\begin{figure}
\centering
\includegraphics[width=0.47\textwidth]{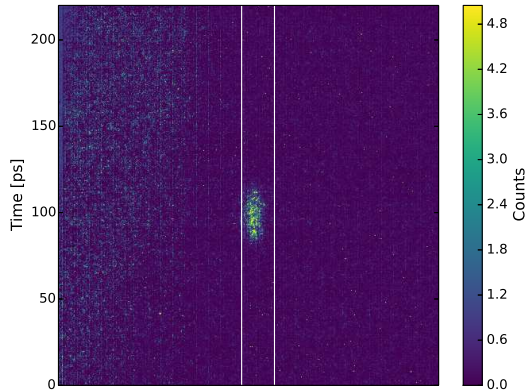}
\caption{Streak-camera image for the up-converted PDC photon, obtained by analog integration over 32 exposures with 10 seconds exposure time each. The white lines indicate the integration boundaries used to obtain the temporal profile. The horizontal axis covers merely the spatial degree of freedom and carries no physical information in this measurement scenario.}
\label{fig:streak1}
\end{figure}

\begin{figure}
\centering
\includegraphics[width=0.44\textwidth]{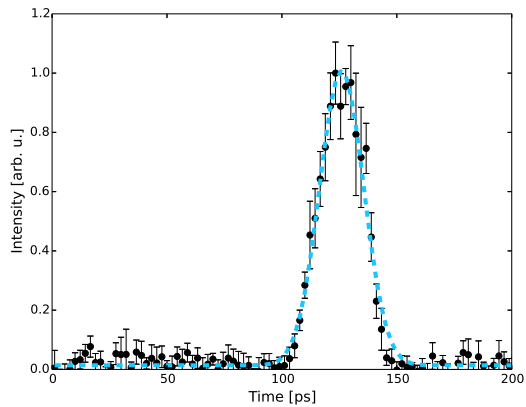}
\caption{Integrated temporal profile of the up-converted PDC photon obtained from the streak-camera image in Figure \ref{fig:streak1}. Error bars were calculated using the standard deviation around the the averaged counts in 5\,ps bins. The blue dashed line indicates a Gaussian fit used to obtain the temporal duration.}
\label{fig:streak2}
\end{figure}

In figure \ref{fig:streak1} we show the resulting background-subtracted image captured with the streak-camera's CCD. The image is integrated over 32 individual exposures. For the background image the PDC beam path was blocked. The multi-channel plate gain was set to two thirds of the maximum value. At larger amplification the cathode noise became the dominating noise source. With these settings the signal was just above the noise floor of the camera, where the readout noise dominated over the dark current. For this reason the largest possible exposure time of ten seconds was chosen for this measurement. Figure \ref{fig:streak2} shows the temporal envelope extracted from Figure \ref{fig:streak1} by integrating in horizontal direction over the area bounded by the white lines.
Error bars were created from the fluctuations around the mean values inside 5\,ps bins. This corresponds to the temporal resolution of the device as given by focus spot size on the photo cathode. Such an approach is reasonable as the imaged spot size on the CCD is much larger than the pixel size.

From previous measurements on the initial temporal envelope in Ref. \cite{allgaier_fast_2017} the pulse duration before the conversion process is known to be 1.1\,ps (FWHM). This up-conversion process alters the temporal envelope and stretches the pulse duration. From theory we expect a duration of 27\,ps long top-hat function. We will elaborate more on this in section \ref{sec:4}. The output is then convolved with the streak-camera's response function given by its limited resolution of 5\,ps. We therefore expect to measure a final pulse duration of 28.5\,ps. However, the measured pulse duration is merely 23\,ps. This is a hint that imperfections in the periodic poling of waveguide structure in our sample yields a shorter effective interaction length as expected. The temporal duration is in fact directly proportional to the interaction length. This observed shorter interaction length is in accordance with the asymmetric phase-matching functions measured in \cite{allgaier_highly_2017}, as a perfectly homogeneous non-linearity profile over the whole crystal length would yield a symmetric sinc phase-matching spectrum, which is not observed in the sample employed here.

\section{Discussion}
\label{sec:4}

There are two key figures that estimate how much the up-conversion scheme improves the overall detection efficiency.  First, the external conversion efficiency of the process, which includes how much light inside the waveguide is converted as well as linear losses, is 27.1\,\% \cite{allgaier_highly_2017}. This linear loss is more than balanced out by the fact that the photo cathode's quantum efficiency is about 3 orders of magnitude higher at 550\,nm than it is at 1550\,nm. The two numbers multiplied give rise to an improvement of the detection efficiency by a factor of 270. Other photo cathode models exist that provide better quantum efficiency at 550\,nm. These photo cathodes have an efficiency enhanced by two orders of magnitude, but may not be sensitive to telecom light at all.
In terms of detection efficiency, our work conclusively shows that up-conversion schemes are viable to make single photons in the infrared range accessible by streak-cameras.

The up-conversion process in combination with the particular Streak camera used in this work facilitate the characterization of telecom photons. However, the configuration of the conversion, in particular the engineered phase-matching, has both advantages and drawbacks. We define the phasemismatch in the traditional way using the wavenumber \(k=2\pi n(\lambda)/\lambda\), where \(n\) denotes the effective refractive index of the waveguide mode, and the poling period \(\Lambda\) used to achieve quasi-phase-matching :

\begin{equation}
\Delta k = k_{\mathrm{pump}} + k_{\mathrm{input}} - k_{\mathrm{output}} + \frac{2\pi}{\Lambda}
\end{equation}

This phase-matching function is the governing quantity of the up-conversion process. Its orientation depends strongly on the dispersive properties of the material and waveguide parameters. In the process employed in this work the phase-matching is flat, which means that the output is mostly independent of the input \cite{brecht_quantum_2011} and there are no spectral correlations between the two.
On the one hand side this provides excellent conversion efficiency and bandwidth compression due to the group-velocity matching between input and pump. 
However, the phase-matching of the process masks the temporal-spectral information of the PDC state. The calculated temporal amplitude of our process at the output of the waveguide reads \cite{baronavski_analysis_1993}

\begin{eqnarray}
s(t) \propto   \int d \omega e^{j \omega t} \int d \omega_2 F_1(\omega-\omega_2)F_2(\omega_2) \nonumber \\
e^{-j \omega_2 \tau} e^{j(\Delta k + \alpha \omega)L/2} \cdot \mathrm{sinc}\left( (\Delta k + \alpha \omega) \frac{L}{2} \right)
\end{eqnarray}

\noindent where \(\alpha=\dot{k}_s-\dot{k}_1\) denotes mismatch between inverse group velocities of output and the two inputs and \(\Delta k\) again denotes the corresponding phase-mismatch.  \(F_1(\omega),F_2(\omega)\) are the spectral intensities of input and pump in respect to angular frequencies, respectively. \(L\) is the sample length. First, the output is defined by a convolution of the pump and input spectra (see integral over \(\omega_2\)). In our case these have comparable shape and bandwidth. Inside the Fourier transform (see integral over \(\omega\)) this convolution is multiplied by a sinc-function which, in the case engineered here, only depends on phase-mismatch and crystal length, as the group velocity mismatch \(\alpha\) is zero. The temporal profile is therefore the Fourier transform of the convolved spectra of input and pump, multiplied with the Fourier transform of the sinc-shaped phase-matching. In the temporal domain, the intensity is therefore the convolution of individual Fourier transforms:

\begin{eqnarray}
S(L,t) = FT\left( \int d \omega_2 F_1(\omega-\omega_2)F_2(\omega_2) \right) \nonumber \\
\ast FT\left( \mathrm{sinc}\left((\Delta k + \alpha \omega) \frac{L}{2} \right) \right)
\end{eqnarray}

We find two competing timescales: One set by the duration of the input fields and one by the crystal length. In our case, the temporal profile is dominated by the contribution of the sinc-function given by the long crystal. The Fourier transform of the spectral overlap integral, which has a duration of the order of 1\,ps, is therefore convolved by a rectangular function of 27\,ps width, which masks the temporal information about the PDC state significantly. It is therefore appropriate to consider alternative process conditions regarding group velocity matching and phase-matching. To achieve a one-to-one mapping between input and output, thus preserving the temporal profile, the phase-matching needs to be at an angle. This is achieved in a group-velocity mismatched process. Here, a small mismatch is sufficient, which leaves the pulse walk-off in the same order of magnitude as the pulse duration. This guarantees that the expected drop in conversion efficiency is still acceptable.

\begin{figure}
\centering
\includegraphics[width=0.44\textwidth]{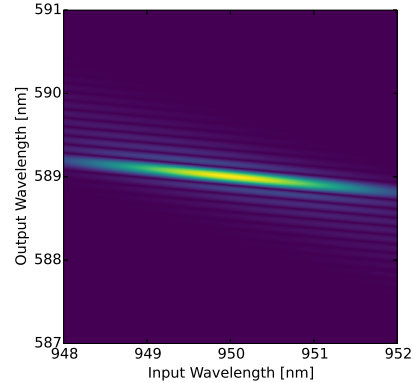}
\caption{Spectral transfer function of the proposed sum-frequency generation process. The angled sinc-shaped phase-matching function is cut at -45 degrees by the pump spectrum to obtain this two-photon intensity that maps input to output wavelength.}
\label{fig:pm}
\end{figure}

We therefore propose a type-0 SFG process in order to realize a more practical up-conversion streak-camera. Such processes were already used in up-conversion detection schemes \cite{pelc_long-wavelength-pumped_2011}. In the process we propose all three fields would be in the extraordinary polarization thus making use of the \(d_{33}\), the highest of the non-linear coefficients of Lithium Niobate. The phase-matching displayed in figure \ref{fig:pm} is at 17\(^{\circ}\) degrees in frequency space, thus providing a bandwidth compression factor of \(1/\tan 17^{\circ} = 3\), therefore stretching the pulse in time by a factor of three. This also implies an improvement on the streak-Camera's resolution from 5\,ps to about 2\,ps, as long as the pump light is pulsed. Contrary to the process employed in this work, the proposed process provides real stretching of the pulse shape instead of reshaping. By exchanging the roles of input and pump the process can be used to convert inputs from a wide spectral range.

In this work the brightness of the converted light was just barely above the detection limit of the streak-camera, however moving from the employed Hamamatsu S1 photo cathode to a at this wavelength more efficient photo cathode would provide an improvement of 100 to the quantum efficiency, and much more to the signal-to-noise ratio due to lower cathode noise.

\section{Conclusion}

We demonstrated the feasibility of using a highly efficient sum-frequency generation in Lithium Niobate to detect single photons at telecom wavelength on a streak-camera. The exceptionally high efficiency of the process makes it feasible to employ streak-cameras for various experiments in quantum optics not only as a means of direct temporal observation of single photons but also as a picosecond-resolution photon counter. For typical fluorescence spectroscopy applications the up-conversion scheme could be used in a variety of ways, for example the characterization of the emission of infrared semiconductor laser structures far below threshold.

\section{Funding Information}

This work was funded by the Deutsche Forschungsgemeinschaft via SFB TRR 142 (C01) (grant number TRR142/1) and via the
Gottfried Wilhelm Leibniz-Preis (grant number SI1115/3-1).

\end{document}